\documentclass[11pt]{article}

\usepackage{graphicx}

\begin{document}

\begin{center}

{\bf\LARGE Comparison of Gluon Bundle based QCD\\ [5mm] Curves with ISR Elastic pp Scattering Data }
\\
\vspace{4cm}  

H.M. FRIED
\\
{\em Department of Physics \\
Brown University \\
Providence R.I. 02912 USA}\\
fried@het.brown.edu\\
[5mm]
P.H. TSANG
\\
{\em Department of Physics \\
Brown University \\
Providence R.I. 02912 USA}\\
peter\_tsang@brown.edu\\
[5mm]
Y. GABELLINI
\\
{\em Institut Non Lin\'eaire de Nice\\
UMR 7335 CNRS\\
06560 Valbonne France}\\
yves.gabellini@unice.fr\\
[5mm]
T. GRANDOU
\\
{\em Institut Non Lin\'eaire de Nice\\
UMR 7335 CNRS\\
06560 Valbonne France}\\
thierry.grandou@inln.cnrs.fr\\
[5mm]
Y.M. SHEU
\\
{\em Department of Physics \\
Brown University \\
Providence R.I. 02912 USA}\\
ymsheu@alumni.brown.edu

\newpage

\vskip1truecm {Abstract}
\end{center}

\indent Using previously described functional techniques for exact, non-perturbative, gauge-invariant renormalized QCD processes, a simplified version of the amplitudes - in which forms akin to Pomerons naturally appear - provides fits to ISR Elastic pp scattering data. Extension of this work to LHC reactions is presently underway. 

{\section{Introduction}}

In a set of recent papers \cite{qcd1,qcd2,qcd3,qcd4,qcd5,qcd6,qcd_summary,qcd7,qcd8} the present authors have shown how it is possible to proceed from any relativistic, gauge-dependent Schwinger generating functional of QCD, to explicit solutions for its correlation functions, in a gauge invariant and non-perturbative manner. This analysis includes the derivation of quark binding and nucleon binding potentials. The present paper describes the application of a simple method of non-perturbative QCD renormalization, in order to proceed from those correlation functions describing hadron scattering to S-matrix elements for high energy, proton--proton scattering. This procedure involves the introduction of a few parameters determined in comparison with the Intersecting Storage Rings, ISR, experimental data \cite{isr_data1,isr_data2,isr_data3,isr_data4}; but once they are fixed, they should be applicable to all QCD scattering reactions at similar energies.

In Section 2, we recall the main features of our non-perturbative approach to QCD, namely the ``effective locality" property, the existence of ``gluon-bundles" in place of ordinary gluons, and its renormalization definition. These will be qualitatively discussed, as the quantitative elements are in the references cited above \cite{qcd1}-\cite{qcd8}.

In Section 3, we set the theoretical frame in which we will do our computation, and set the gluon-bundle renormalization conditions, specific for the calculation of this pp scattering cross section.

In Section 4, we give our  ``postdiction" of $\displaystyle\frac{d\sigma}{dt}$ to be compared to the ISR data, with some insight of how to go from our QCD formula to a computationally simpler expression.

Section 5 is the Summary, in which we discuss slight generalizations of the methods used in this paper to account for pp scattering at LHC and even higher  energies. We also comment on the form of total cross-section predicted by our analyses, and the criterium needed to insure an absence of violation of the Froissart bound.

\begin{figure}
\centering
\includegraphics[width=12cm]{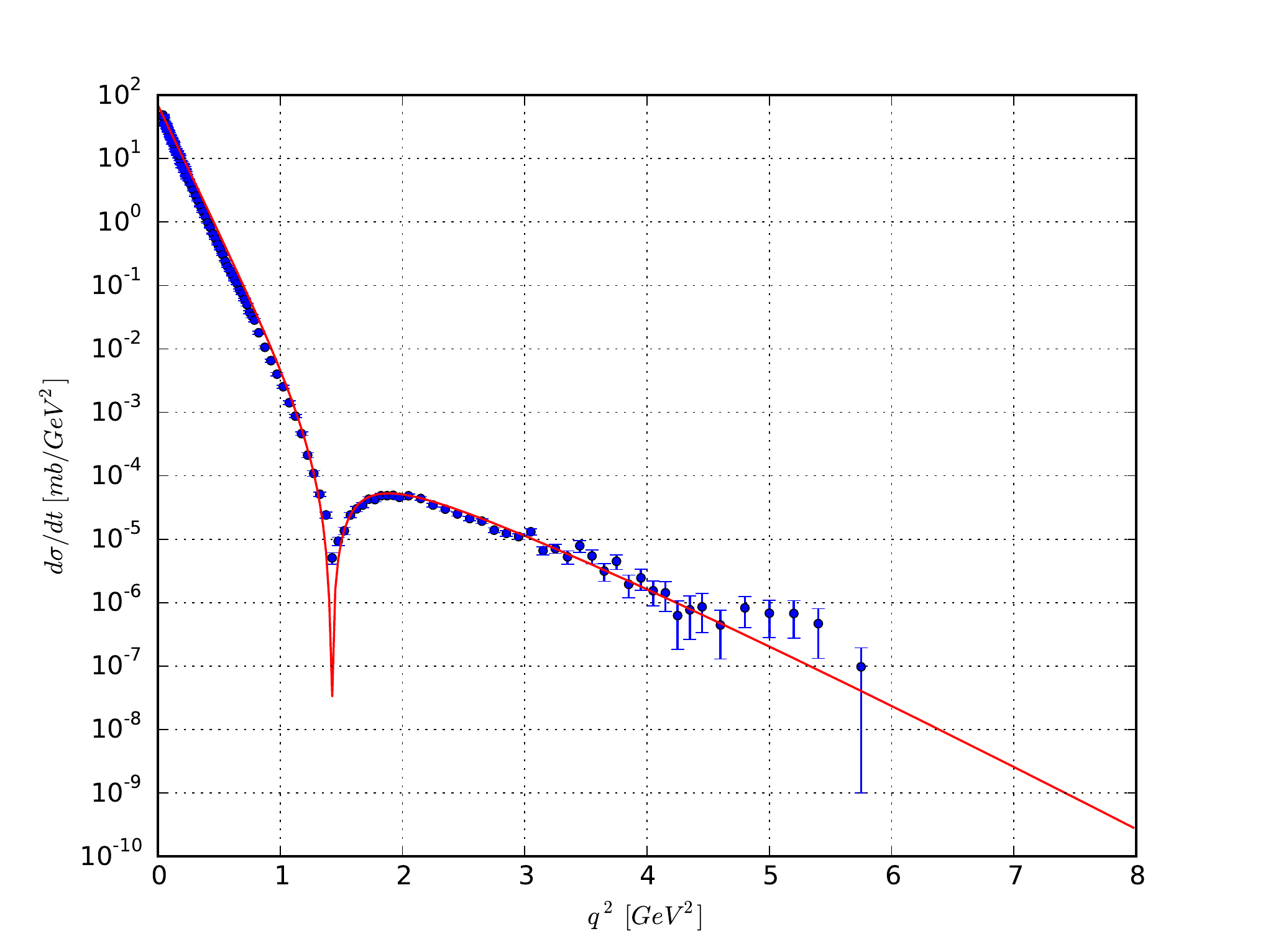}
\caption{Elastic pp scattering differential cross section at $\sqrt{s} = 23.5$ GeV. Blue dots are experimental data, red line is the result from Eq.(\ref{eq:1}) }
\label{24gev}       
\end{figure}

\begin{figure}
\centering
\includegraphics[width=12cm]{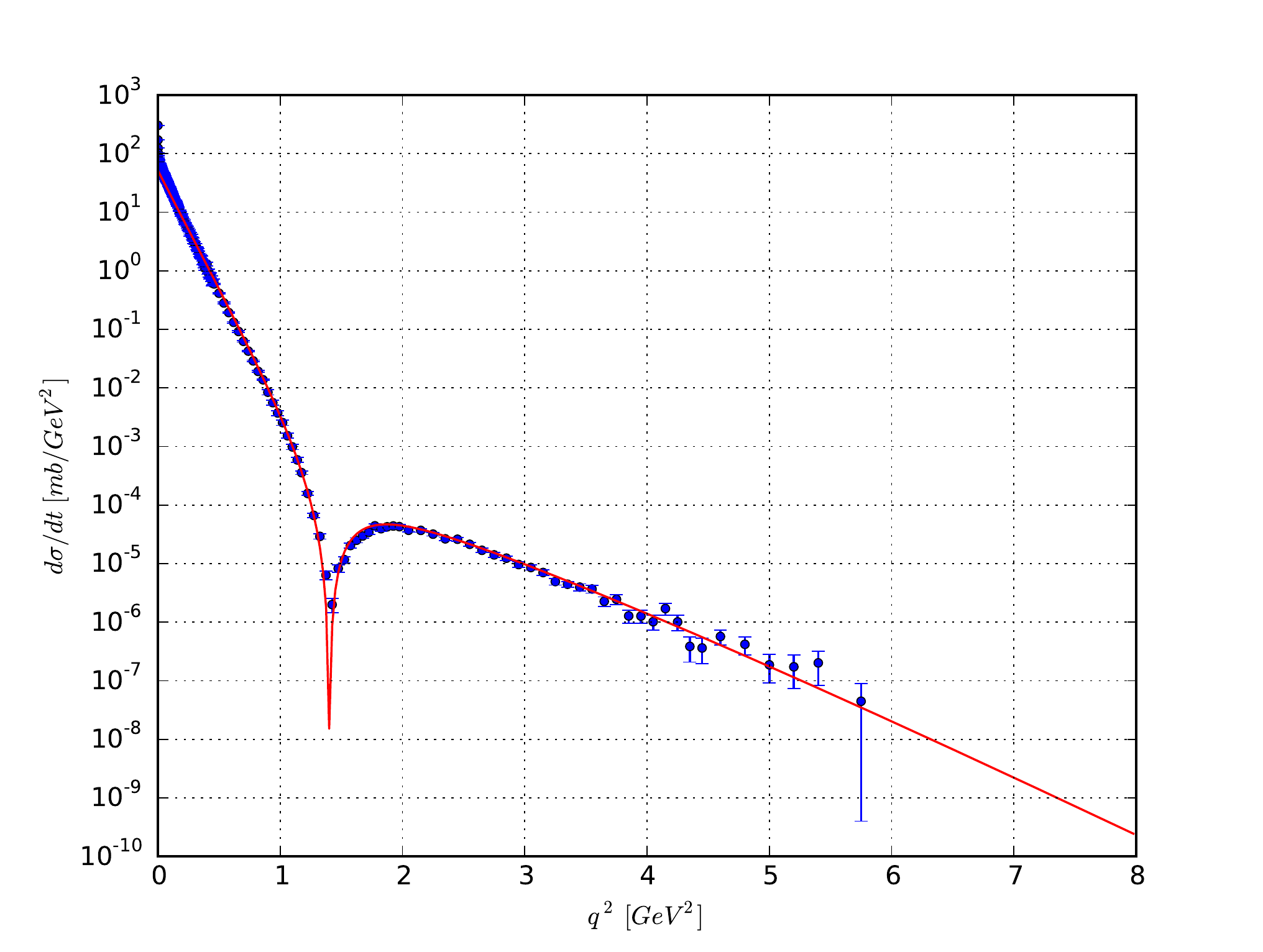}
\caption{Elastic pp scattering differential cross section at $\sqrt{s} = 30.7$ GeV. Blue dots are experimental data, red line is the result from Eq.(\ref{eq:1})}
\label{31gev}
\end{figure}

\begin{figure}
\centering
\includegraphics[width=12cm]{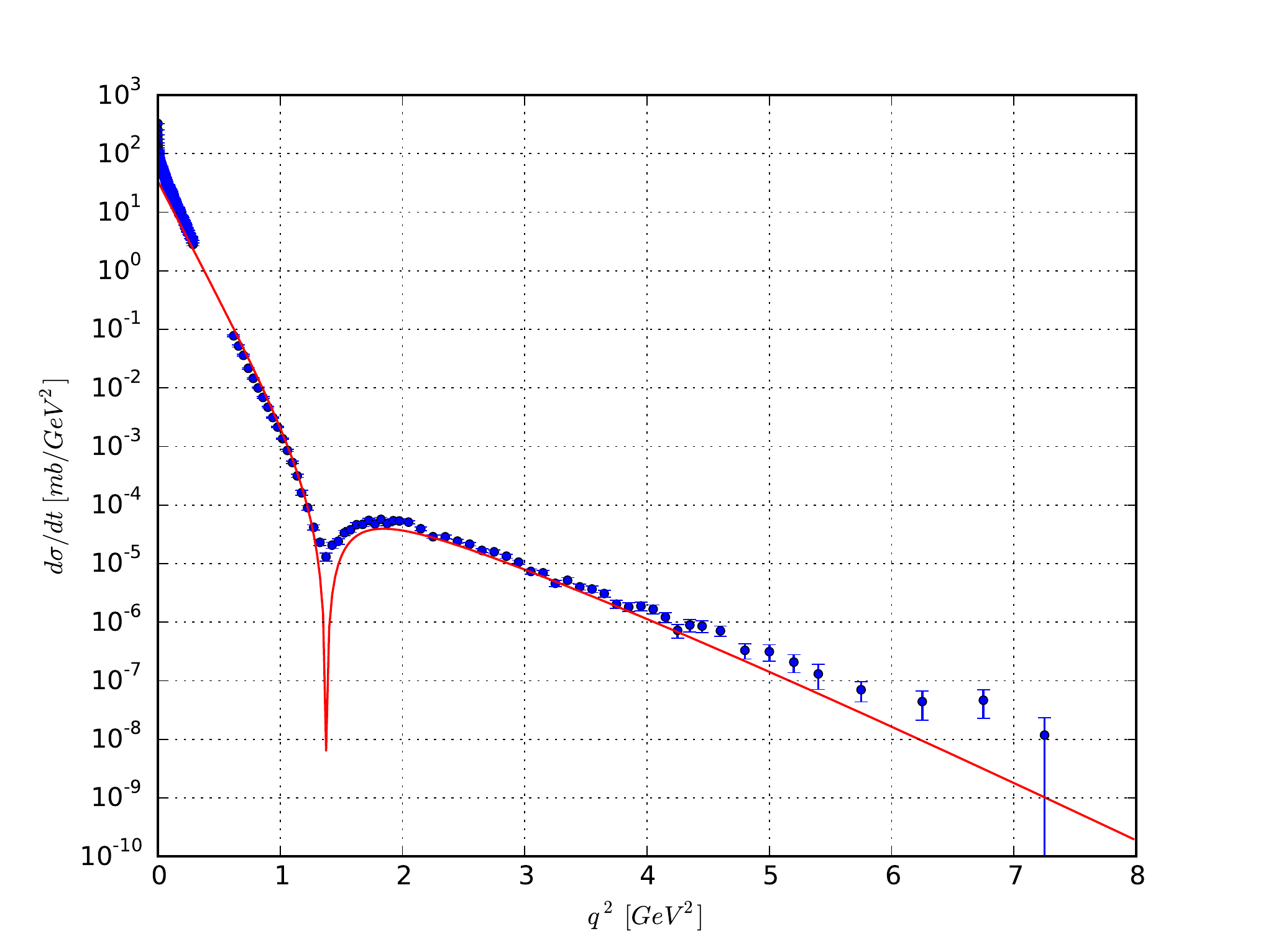}
\caption{Elastic pp scattering differential cross section at $\sqrt{s} = 44.7$ GeV. Blue dots are experimental data, red line is the result from Eq.(\ref{eq:1})}
\label{45gev}       
\end{figure}

\begin{figure}
\centering
\includegraphics[width=12cm]{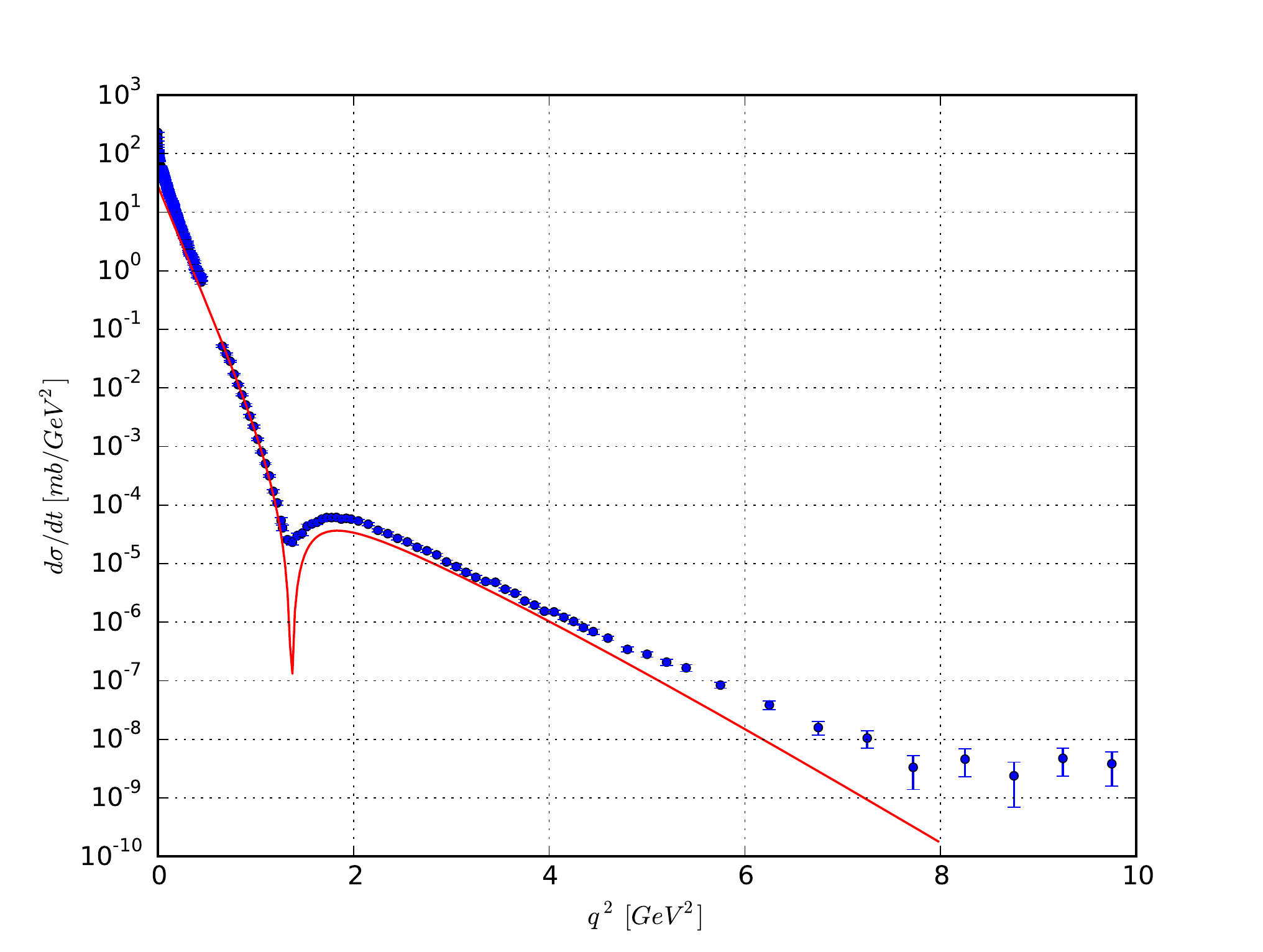}
\caption{Elastic pp scattering differential cross section at $\sqrt{s} = 52.8$ GeV. Blue dots are experimental data, red line is the result from Eq.(\ref{eq:1})}
\label{53gev}       
\end{figure}

\begin{figure}
\centering
\includegraphics[width=12cm]{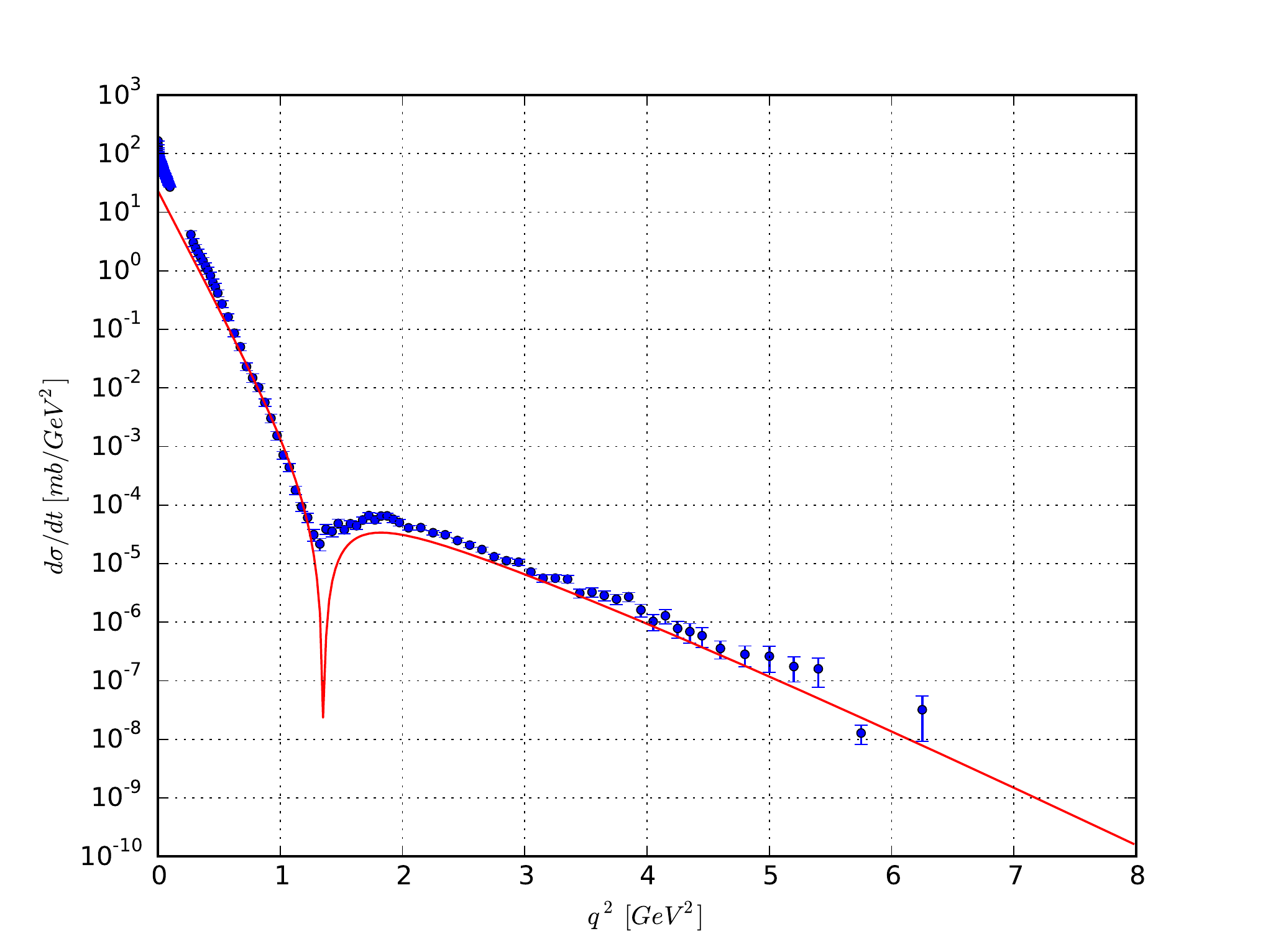}
\caption{Elastic pp scattering differential cross section at $\sqrt{s} = 62.5$ GeV. Blue dots are experimental data, red line is the result from Eq.(\ref{eq:1})}
\label{63gev}      
\end{figure}

In any paper on High-Energy diffractive scattering, one should first make reference to the very well-known Pomeron formulation, of which we simply quote two detailed and thorough references \cite{pomeron,diffraction}.  And as can be seen from the fits of Figures \ref{24gev}-\ref{63gev}, we have, in effect, given field theory derivations of what might be called the non-perturbative QCD Pomerons.

\bigskip
{\section{Non Perturbative QCD}}

Let us start by mentioning the general thrust of our approach, which is to begin with the Schwinger generating functional, with gluons in any relativistic gauge; and then perform a simple rearrangement which brings the generating functional into a completely gauge-invariant form. At that stage, the exact functional operations required by the generating functional are made possible by the use of two Fradkin functionals, gaussian in their field variable \cite{fradkin1,fradkin2}, together with a familiar gaussian relation written and used by Halpern \cite{halpern1,halpern2}; and the result of this functional operation is then the appearance of a new and exact property of non-perturbative QCD, called ``Effective Locality" (EL) ~\cite{qcd2,grandou_casimir1,grandou_casimir2}, and the explicit demonstration of the gauge-independence of the theory. The immediate consequence of EL is the replacement of a main functional integral by a set of ordinary Lebesgue integrals, which can be evaluated exactly but are easily estimated using pencil and paper, or by a laptop for more accurate results. 

As derived in our QCD papers, one finds that all radiative corrections to the correlation functions of non-perturbative, gauge independent QCD are obtained by the exchange of ``Gluon Bundles" (GBs) between any pair of quarks and/or antiquarks, including quarks which form virtual, closed quark loops, and those which are, or are about to be bound into hadrons. Each GB consists of a sum over an infinite number of gluons, with space-time and color indices properly maintained and displayed. One is then able to define quark-binding potentials without the use of static quarks, and to produce a qualitative nucleon-binding potential, in which two nucleons form a model deuteron \cite{qcd4,qcd6}.

	One very important point to notice is that individual gluons do not appear in this formulation, for they have all been summed and incorporated into GBs. The functional formalism we use makes the extraction and summation of such effects a relatively simple matter, although the correlation results can still be quite complicated. 

\begin{figure}
\centering
\includegraphics[width=12cm]{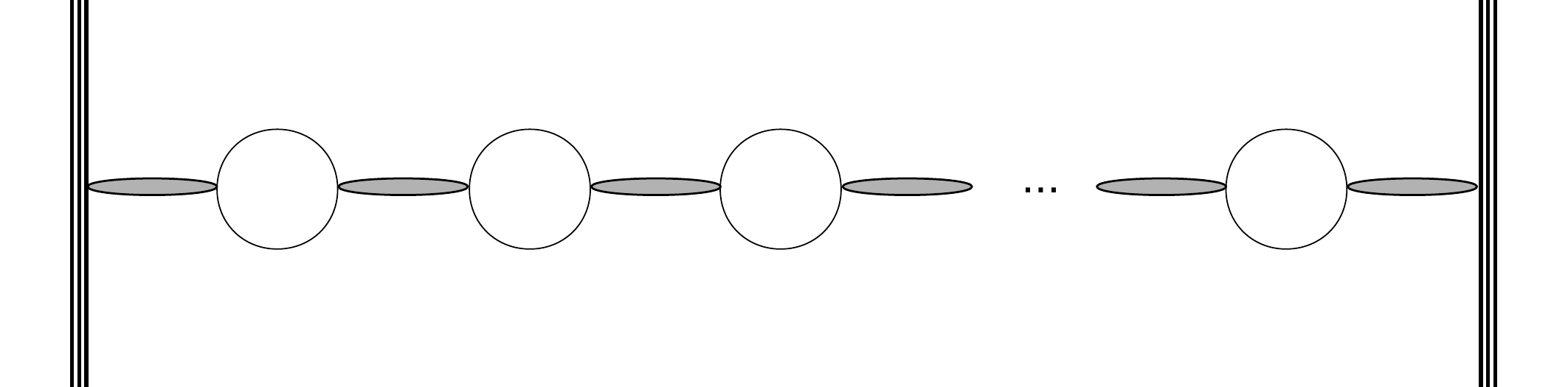}
\caption{Renormalization adopted such that no more than two gluon-bundles are attached to each quark loop; and each chain ends on a quark bound into a hadron.}
\label{chainloop}      
\end{figure}

	Concerning renormalization, for a GB there is no hint, no previous problem to which one can turn for intuitive assistance; rather the question of GB renormalization may, in part, be decided by subsequent simplicity, and with the parameters of that renormalization fixed by the data. That passage from correlation functions to S-matrix elements was described in our most recent paper \cite{qcd8}, in which non-perturbative QCD renormalization was defined. In this formulation there can no longer be any reference to individual gluons, and conventional perturbative renormalization must be redefined in terms of GBs interacting with quark loops, and with quarks forming hadronic bound states. A special and surely the simplest form of renormalization was adopted, in which quark loops automatically appear only in chains, with no more than two GBs attached to each loop; and each chain ends on a quark bound into a hadron, as in Fig.\ref{chainloop}.

The calculations can in principle all be defined and carried through exactly, and in a finite manner; but for simplicity and ease of presentation, certain obvious approximations were presented. These simplifications are retained in the present paper, in which the above analysis is applied to the ISR elastic scattering of two protons, at a variety of energies in the GeV range \cite{peterthesis}. There will appear below an additional set of simple approximations to specific integrals, again for reasons of subsequent simplicity. 
\bigskip

{\section{Elastic pp Scattering and Gluon Bundle Renormalization}}

\begin{figure}
\centering
\includegraphics[width=10cm]{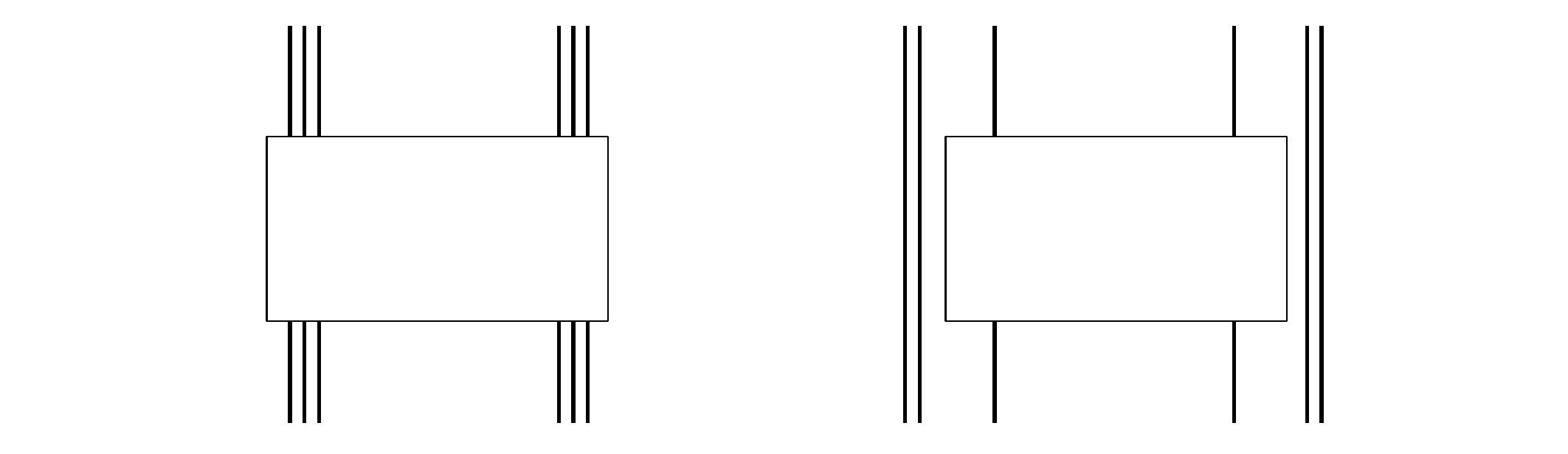}
\caption{The elastic pp scattering. On the left, the six-body interaction; on the right, the two-body approximation.}
\label{ppscat}       
\end{figure}

	We emphasize that these descriptions of pp scattering can, in principle, be evaluated exactly in terms of six-body quark interactions, using Random Matrix methods \cite{qcd7,grandou_casimir2,random_matrices}, but in order to keep this paper one of finite length, we have employed several approximations when evaluating relevant integrals. Perhaps the most serious simplification has been performed at the very beginning, by assuming that the scattering is "truly elastic", so that each triad of scattering quarks remain bound into its initial proton during the entire scattering process. This precludes, for example, the interchange of any quarks comprising each proton, as well as other more complicated possibilities, and is clearly incorrect as energies increase. But it does replace a six-body quark problem by a two-body scattering problem (Fig.\ref{ppscat}); and the corrections to this two-body approximation are easily and intuitively defined, by the insertion of a weak energy dependence, phenomenologically obtained from the data. While it is important to understand that the correlation functions of our QCD functional procedure can be exactly calculated, it is surely a computational and physical advantage to employ the two-body approximation, which is almost but not quite true at ISR energies. And it will be less true, but still useful, at LHC energies.

	We now turn to a detailed treatment of GB renormalization, specific to the present problem, in which we imagine each proton as a bound state of three quarks, with these three quarks here interacting with the three quarks of the other proton. It should be understood that even at such low energies we are completely neglecting electrodynamic effects, and quark spin effects, which can always be added separately.
	From the original definition of the Halpern functional integral, plus the appearance of Effective Locality, at each end of a Gluon Bundle there appears a quantity $\delta$, which divides into two classes: those which connect to a quark which is, or is about to be bound into a hadron; and those which connect to a quark loop. Before renormalization, each of these $\delta$ must vanish; but renormalization here means that: 

	1) the $\delta$ at the quark loop end of the gluon-bundle is to vanish. Combined with the expected UV log divergence of the loop, $\ell$, this gives a finite --and small-- parameter: $\kappa = \delta^2 \ell$, which we take as a fixed, real, positive parameter, extracted from the data.
	2) But this is not true for quarks of the first group, for such quarks are  the "physical particle" of QCD where $\delta$ becomes a finite quantity $\delta_q(E)$.  Each $\delta_q(E)$ has a dimension, which we may think of as time, or distance; and thanks to the Heisenberg inequality, the natural choice is to replace that $\delta_q$ by a dimensionless constant multiplying $1/E$, even though this leads to a rapid decrease of the differential cross section as the energy increases. But as the energy increases to ISR values, one finds that cross section for all ISR scatterings is about the same, although still decreasing, but very slowly. The reason is presumably the onset and continued growth of  "quasi-state" processes: there are more and more ways of interchanging quarks and combining quarks and loops to produce a final state of two protons. The shape of the $q^2$ dependence of the ISR scatterings is barely affected, and this is presumably due to the fact that however complicated the intermediate “quasi-states" might be, the end product of each elastic process must be two protons.

	Following this interpretation, we must now change to a specific form of $\delta_q(E)$, one which permits a very slow decrease with increasing energy; and for this we have chosen $\delta_q(E) = (1/m)(m/E)^p$, where $0\!<\!p\!<\!1$, to be chosen by the data. Of course, this is a phenomenological choice of the variation with energy of all amplitudes so constructed in the ISR range, and seems to be the best one can do under the two-body restrictions. 

\bigskip
{\section{Evaluating the ISR Postdictions}}

We here present a qualitative description of the essential steps of our derivation, the details of which may be found in Ref.\cite{qcd8}. The first step is to write the eikonal representation of any Quantum Field Theory scattering amplitude, with its associated differential cross section:

\begin{equation}
\matrix{\displaystyle T(s,\vec q)&\displaystyle =\frac{is}{2m^{2}}\int d^{2}b\ e^{\displaystyle i\vec q\cdot \vec b}\ [1-e^{\displaystyle i{\bf X}(s,\vec b)}]
\hfill\cr\noalign{\medskip} \hfill\displaystyle\frac{d\sigma}{dt} & \displaystyle = \frac{m^4}{\pi s^2}\, | T |^2\hfill}
 \end{equation}
where ${\bf X}(s,\vec b)$ is the eikonal function appropriate to the scattering, when $s=4E^2$, where $E$ is the center of mass energy of each incident proton, and $\vec q$ is the momentum transfer in that frame: $\vec q^{\,2} = |t |<< s$, $m$ being the mass of that proton.

That eikonal is constructed in terms of multiple gluon-bundles and closed loop chains between each quark of each triad of the three quarks defining each asymptotic proton (with appropriate and hidden statements of the binding of each triad, which are to be understood); and as derived in Ref.\cite{qcd8}, this amplitude may be expressed in terms of the relevant gluon-bundles and loop chains exchanged, in multiple ways, between the quarks of different protons. And finally, for the ISR amplitudes, which are decreasing with increasing energy, it is appropriate to expand that eikonal, retaining only its 2 gluon-bundle portion, plus the geometric series of all odd-loop chains; other terms in that expansion would produce correspondingly smaller corrections in powers of $(m/E)^p$.
\begin{figure}
\centering
\includegraphics[width=10cm]{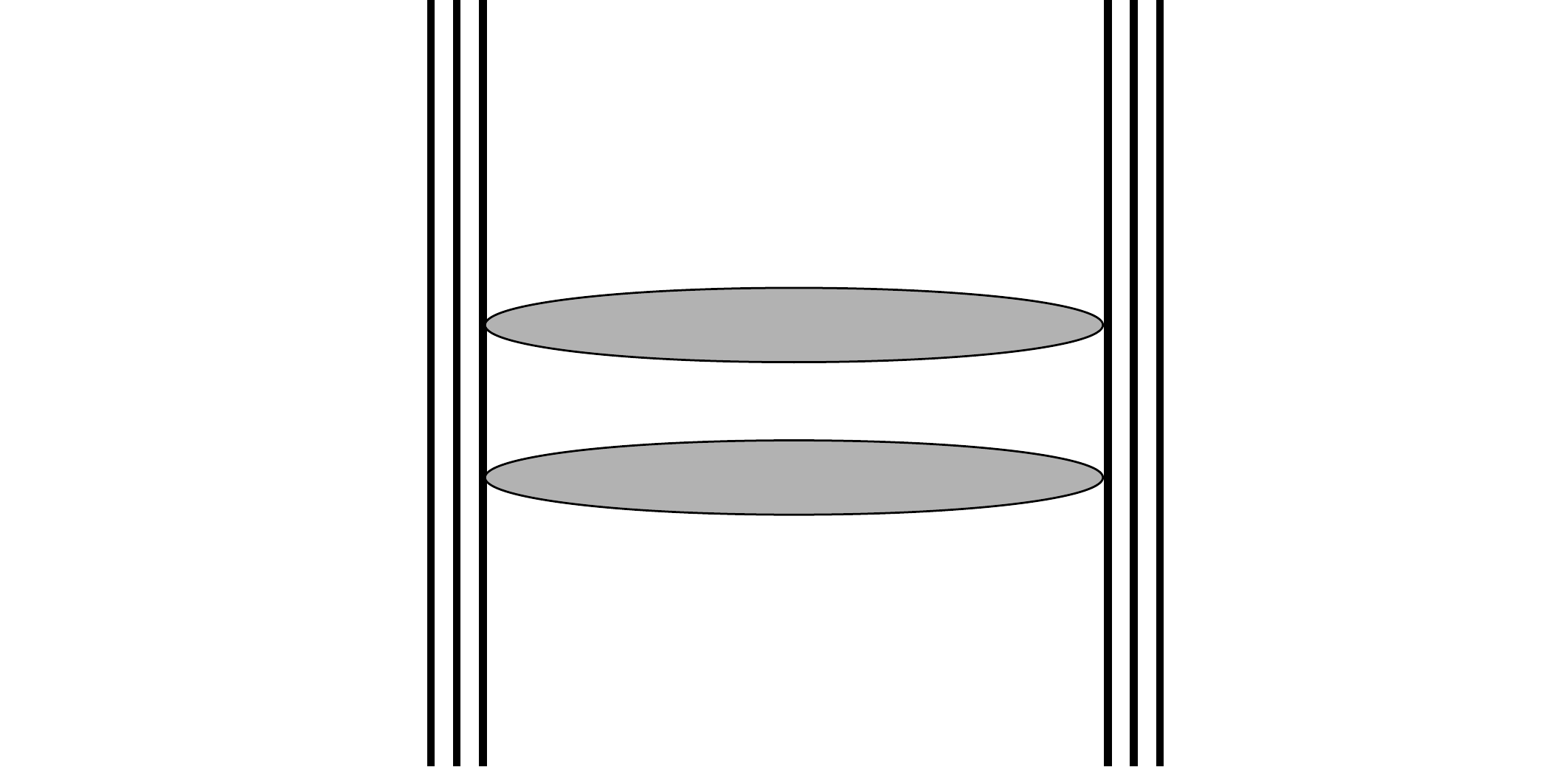}
\caption{The two gluon-bundle exchange as the first ingredient for the ISR amplitude (left part of the bracket of Eq.(\ref{eq:1})).}
\label{gluonbundle}       
\end{figure}

\begin{figure}
\centering
\includegraphics[width=10cm]{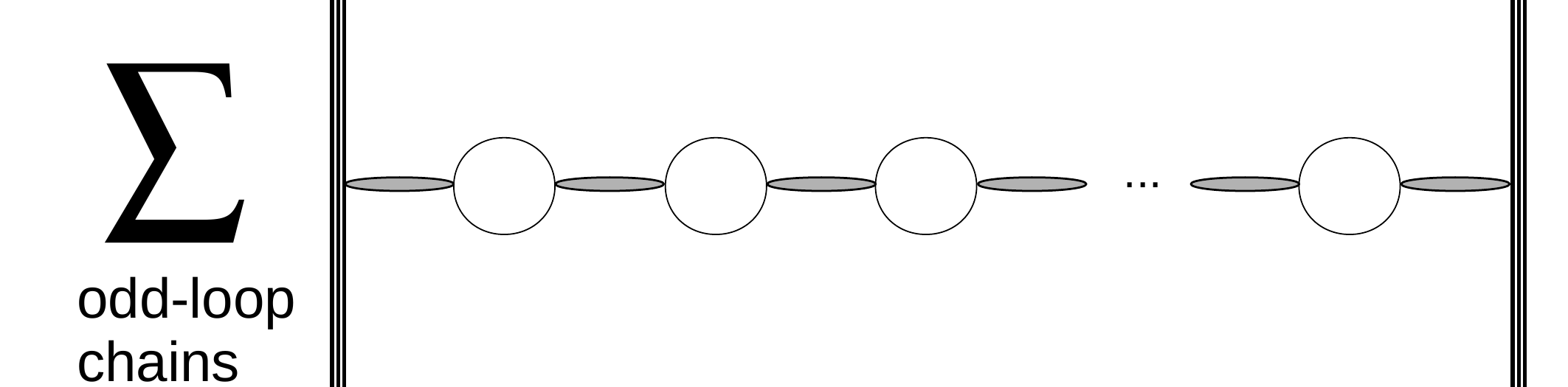}
\caption{The sum of all odd-loop chains as the second ingredient for the ISR amplitude (right part of the bracket of Eq.(\ref{eq:1})).}
\label{oddloops}       
\end{figure}

We now make two additional approximations in the evaluation of integrals such as Eqs.(13) or (31) of Ref.\cite{qcd8}, and those that follow in subsequent paragraphs. The first simplification is to replace the exact integral:
\begin{equation}
e^{\displaystyle i{\bf X}(s,\vec b)} = N \int d[\chi] e^{\displaystyle i/4 \int \chi^2}[\det(f\!\cdot\!\chi)^{-1}]^{1/2}\ \mathcal{F}\big(k'(f\!\cdot\!\chi)^{-1}+iC(f\!\cdot\!\chi)^{-2}\big)
\end{equation}
derived in Ref.\cite{qcd8} from the original Halpern functional integral, where $\mathcal{F}$ represents the exponential of a gluon-bundle exchanged between two quarks of different protons, plus an entire loop chain connecting those quarks, where $f\!\cdot\!\chi = f^{abc}\chi^c_{\mu\nu}$, $f$ are the antisymmetric SU(3) structure constants and $\chi$ -- to be integrated over -- is antisymmetric in its Lorentz indices. $((gf\!\cdot\!\chi)^{-1})^{ab}_{\mu\nu}$ is the quantity caracterizing the gluon-bundle and its locality property: $<\!x|(gf\!\cdot\!\chi)^{-1}|y\!> = (gf\!\cdot\!\chi(x))^{-1}\,\delta^{(4)}(x-y)$.  Eq.(2) can be expressed in calculable form by the introduction of Random Matrix Methods ~\cite{qcd7,grandou_casimir2,random_matrices}, but a definite simplification will be obtained by replacing $f\!\cdot\!\chi$ by $R$, where $R^2$ denotes the magnitude of $(f\!\cdot\!\chi)^2$, and all of its color-angular integrations are supressed. This simplification assumes that the color-angular integrations over different color coordinates have no real bearing on the dynamical outcome of (2); and that the important part of the exact (2) will depend only on the magnitudes of $f\!\cdot\!\chi$. It should also be mentioned that $(f\!\cdot\!\chi)^{ab}_{\mu\nu}$, as well as its inverse, is antisymmetric under the full $\int d[\chi]$ and that therefore only even powers of $R$ will be retained.

That is, we rewrite (2) in the form:

\begin{equation}
e^{\displaystyle i{\bf X}(s,\vec b)} =  N'\int_0^{\infty} \!dR\,R^7 e^{\displaystyle i\beta R^2}\,R^{-4}\,\mathcal{F}(R)
\end{equation}
where the determinant factor of (2) has been replaced by the $R^{-4}$ of (3), $\beta$ is a real constant, to be defined from the data,  $\mathcal{F}(R) = \mathcal{F}\big(k'R^{-1}+iCR^{-2}\big)$ and $N'$ is the new normalization constant such that for $g=0$, $\mathcal{F}=1$:

\begin{equation}
  N'\int_0^{\infty} \!dR \,R^3\ e^{\displaystyle i\beta R^2}=1
  \end{equation}
  
The second simplification performed for these ISR amplitudes is in their evaluation of the integration over the $R$ magnitudes. The integral of Eq.(4) needs no simplification, and can be performed directly, yielding $N'=-2\beta^2$. But for the more complicated integrals, of the form of (3), we use the following scheme to approximate (3):

\begin{equation}
e^{\displaystyle i{\bf X}(s,\vec b)} =  N'\int_0^{\infty} \!dR\,R^3 e^{\displaystyle i\beta R^2}\mathcal{F}(R) = N'\Big(-i \frac{\partial}{\partial \beta}\Big)\int_0^{\infty}\!dR\,R \  e^{\displaystyle i\beta R^2}\ \mathcal{F}(R)
\end{equation}
and with the variable change:

\begin{equation}
R^2=iu\ ,\ \ \  R = \sqrt{iu} = e^{i\pi/4}\sqrt{u}
\end{equation}
one obtains:

\begin{equation}
e^{\displaystyle i{\bf X}(s,\vec b)} =  \frac{N'}{2} \Big(\frac{\partial}{\partial \beta}\Big)\int_0^{\infty} du\ e^{\displaystyle -\beta u}\ \mathcal{F}\Big({\sqrt{iu}}\Big)\ .
\end{equation}

The integral of (7) has serious contributions only for $u < 1/\beta$ which we approximate as:

\begin{equation}
e^{\displaystyle i{\bf X}(s,\vec b)} =  \frac{N'}{2} \Big(\frac{\partial}{\partial \beta}\Big)\int_0^{1/\beta} du\ \mathcal{F}\Big({\sqrt{iu}}\Big)= \mathcal{F}\Big(\sqrt{i/\beta}\,\Big)\ .
\end{equation}

We again emphasize that our functional representations can, in principle, all be calculated exactly; but to keep this paper more easily readable, we have resorted to the two approximations of this Section. It should also be noted that the constant $C$ of Eq.(2) follows from an integration over $\int d[\chi(0)]$ of the sum of interior gluon-bundle chain contributions, as defined in Ref.\cite{qcd8}; and that in its evaluation, the two approximations above have been used in exactly the same way, with the same value of $\beta$. In an exact rendering of the integrals of Eq.(2), the parameter $\beta$ would not appear; and we observe here that the same value of $\beta$  - see the list below -  can fit the data, which value turns out to be quite close to the factor of $1/4$ multiplying the exponential of $\chi^2$, in the original Halpern representation. 

This seemingly casual replacement of a truly complicated integral, Eq.(2), has one further advantage: $(f\!\cdot\!\chi)^{ab}_{\mu\nu}$ being an antisymmetric tensor in its color variables, this permits a simplified counting of the number of allowed ways in which gluon-bundles and quark loop chains can be exchanged between different quarks in the other proton.

Because of the smallness of the $\kappa =  \delta^2 \ell$ parameter, most of the effect arises from the 2 gluon-bundles exchange (see Fig.\ref{gluonbundle}) -- a single gluon-bundle exchange would give 0 (see below) -- plus the odd loop contribution of the loop chain, the first with but one loop, followed by the three, five, etc, loop contribution (see Fig.\ref{oddloops}); as noted in Ref.\cite{qcd8}, a term with an even number of loops -- and therefore an odd number of gluon-bundles -- vanishes because the $\int d[\chi]$ parameters are to be integrated from $-\infty$ to $+\infty$. Including all the rest of the odd-numbered loops of that chain, which form a geometric series that can be easily summed, the total contribution exchanged between a quark of one proton and a quark of the other proton, including the multiplicity factors that represent the number of ways in which such quantities can be exchanged, our approximate formula to represent elastic pp scattering at ISR energies is:

\begin{equation}
  \frac{d\sigma}{dt}(E,q^2) = K\bigg[g^2\beta\ \Big(\frac{m_{ext}}{E}\Big)^{2p}\bigg]^2\bigg[ \frac{1}{4\pi}(9\times 3\times 4)\ \Big(\frac{m_{ext}}{E}\Big)^{2p}e^{\displaystyle{-(3q^2/8m_{ext}^2})} \,-\,\frac{(9\times 3\times 8)\ A_{ext}(q^2)}{1+\beta^2g^2A^2_{int}(q^2)}\bigg]^2
 \label{eq:1}      
 \end{equation}
where $A(q^2)=\kappa \,(q^2/m^2)\,e^{\displaystyle{-(q^2/4m^2)}}$, and the $A$ subscripts refer to the $m$ factors used for the ``external" gluon-bundles attached to bound quarks, while the $m$ values of those gluon-bundles attached only between loops are denoted by ``internal" subscripts. Concerning the choice of values for these mass parameters, the inverse of $m$ is proportional to the size of the tranverse gluon fluctuations between different quark lines, which quantities our approximations cannot determine, and their values must here be fixed by comparison with the data. However, the two $m$ values chosen are intuitively clear, with the ``exterior" parameter $m_{ext}$ much closer to a pion mass than is the larger ``interior" $m_{int}$. The gluon-bundle multiplicity factor, $9\times 3 \times 4$, represents the total number of ways of choosing a pair of identical gluon-bundles, and the loop multiplicity factor $9\times 3\times8$ represents the multiplicity possible for the loop chain connections to the equivalent quarks of the different protons. 

We list the values of the fixed parameters of Eq.(\ref{eq:1}):
\medskip

$K=0.44$ mb GeV$^{-2}$

$g=7.6$ 

$\beta = 0.30$

$m_{ext} = 0.28$ GeV $\simeq 2m_{\pi}$

$m_{int} = 0.44$ GeV $\simeq 3m_{\pi}$

$p = 0.14$

$\kappa = 5.22 \,10^{-6}$

We expect that, with the exception of $K$ and $\kappa$, these parameters may have a slight dependence on energy, as it increases from ISR to LHC values, and higher; and that such changes would be due to our two-body approximation of this six-quark scattering reaction. 

\bigskip
{\section{Summary}}

The forms and results of the above calculations and data fits (Fig.~\ref{24gev}-\ref{63gev}) are suggestive of what might happen in other calculations of QCD processes. There may, in fact, be a small but non trivial parameter which effectively defines a possible perturbative sequence: the loop renormalization parameter $\kappa$, which appears to be so small that it might be used to systemically neglect higher numbers of closed quark loops. In the scattering problem, one must retain at least one loop, in order to show a minimum followed by a $q^2$ dependent rise and then fall of the differential cross section with increasing $q^2$. If the energy dependence is too large, this may not work, but it is an interesting point to check; if the coefficients multiplying $\kappa$ are sufficiently small, one then would have a simple and realistic method of approximating a vast number of "non-perturbative" processes.

A comment on the suggested appearance of Pomerons, resulting from our non-perturbative analysis may be appropriate. An immediate statement is that, in no way, are our results specifically related to any of the many perturbative calculations and Reggeon estimations of soft and hard Pomerons; but we do find a natural separation of our amplitudes and differential cross-sections into a dominant part at small momentum transfers, and another part which becomes important at larger momentum transfers. For example, in Fig.~\ref{24gev}, for $q^2$ values less than the dip position at about $1.5\ GeV^2$, that contribution of two GB terms is dominant; while rising from zero, and for $q^2$ values larger than that of the dip, it is the closed-loop-chain which plays the dominant role. If one wishes to use Pomeron terminology, one can refer to these respective contributions as "Non-perturbative Soft and Hard Pomerons".

Finally, a remark on the situation at LHC and higher energies, which are presently under calculation. If our phenomenological representation of the energy dependence inserted during GB renormalization continues in the same way as adopted for the ISR data - although with a different parameter - as energies increase there will come a region in which the expansion of the eikonal - an exponential given by the sum of a GB plus a complete closed-loop-chain - is no longer valid in its simplest, linear form. In effect, one must expect an interference between different (even) numbers of GBs and of loop chains. Initial estimations seem to suggest that the first example of such interference will happen for LHC energies; and that the result will be a "smoothing" of the dip, which by then has moved down to about $.5\ GeV^2$. One can also note that the Total Cross Section depends only upon the sum of all (even) GBs; and that our estimates for this quantity are well below the Froissart bound.

\bigskip
\vskip1truecm 
\noindent{\bf\Large Acknowledgement}
\vskip0.3truecm 
This publication was made possible through the support of a Grant from the Julian Schwinger Foundation. We especially wish to thank Mario Gattobigio for his many kind and informative conversations relevant to the Nuclear Physics aspects of our work. It is also a pleasure to thank Mark Rostollan, of the American University of Paris, for his kind assistance in arranging sites for our collaborative research when in Paris.

\end{document}